\documentstyle[12pt,blois,psfig]{article}

\newcommand{\apg}{\:^{>}_{\sim}\:}

\newcommand{\kms}{\mbox{km\ s${^{-1}}$}}
\newcommand{\lya}{\mbox{${\rm Ly}\alpha$}}
\newcommand{\civ}{\mbox{${\rm C\ IV}$}}
\newcommand{\etal}{et al.}
\newcommand{\hI}{\mbox{${\rm H\ I}$}}

\begin{document}
\heading{THE GASEOUS EXTENT OF GALAXIES AND THE ORIGIN OF QSO ABSORPTION LINE SYSTEMS}

\author{H. -W. Chen $^{1}$, K. M. Lanzetta $^{1}$, J. K. Webb $^{2}$, 
X. Barcons $^{3}$, and A. Fern\'{a}ndez-Soto $^{2}$} {$^{1}$ Department of 
Physics and Astronomy, State University of New York, \\
Stony Brook, NY 11794--3800, USA} {$^{2}$ School of Physics, University of 
New South Wales,\\
Sydney 2052, NSW; AUSTRALIA} {$^{3}$ Instituto de F\'\i sica 
de Cantabria (Consejo Superior de Investigaciones \\
Cient\'\i ficas---Universidad de Cantabria), Facultad de Ciencias, 39005 
Santander, SPAIN}

\begin{bloisabstract}
We present results of an ongoing program to study the gaseous extent of 
galaxies and the origin of QSO absorption line systems. For \lya\ absorption
systems, we find that absorption equivalent width depends strongly on galaxy 
impact parameter and galaxy $B$-band luminosity, and that the gaseous extent of
individual galaxies scales with galaxy $B$-band luminosity as 
$r\propto L_B^{0.40\pm0.09}$. Applying the results to galaxies in the Hubble
Deep Field to calculate the predicted number density of \lya\ absorption lines 
as a function of redshift and comparing it with observations, we find 
that at least 50\% and perhaps as much as 100\% of \lya\ absorption systems 
with $W\apg0.32$ \AA\ can be explained by the extended gaseous envelops of 
normal galaxies. The anti-correlation analysis has also been performed on 
\civ\ absorption line systems, and the results show that the ionized gas cross 
section scales with galaxy $B$-band luminosity as $r\propto L_B^{0.76\pm0.26}$.
\end{bloisabstract}

\section{Background}

The absorption line systems observed in the spectra of background QSOs have
provided a unique probe to the distant universe and an unbiased way to study 
the extended gas surrounding intervening galaxies. Comparison of galaxy and 
absorber redshifts along common lines of sight shows that there exists a 
distinct anti-correlation between \lya\ absorption equivalent width $W$ and 
galaxy impact parameter $\rho$ at $z<1$, although with a substantial scatter 
about the mean relation \cite{l95}. This result suggests that \lya\ absorption 
systems arise in extended gaseous envelopes of normal galaxies and that the gas
cross section is dependent on more than galaxy impact parameter.

To investigate how the gas cross section depends on the properties of 
individual galaxies, we have initiated a program to obtain and analyze HST 
WFPC2 images of galaxies identified in fields of HST spectroscopic target QSOs.
We measure structural parameters, angular inclinations and orientations, 
luminosities, and rough morphological types of 117 galaxies in 17 QSO fields 
\cite{c98}. Galaxy and absorber pairs are then formed according to the
comparison of redshifts.

\section{The extent of \hI\ gas surrounding galaxies}

To see how the gas cross section depends on galaxies of different properties, 
we show in Fig. 1 the residuals of the $W$ vs. $\rho$ anti-correlation, defined
to be $\log W - \alpha\log \rho - C$, where $\alpha$ and $C$ are determined 
from a maximum likelihood analysis, as a function of galaxy $B$-band absolute 
magnitude $M_B$, redshift $z$, mean surface brightness $\langle\mu\rangle$, and
disk-to-bulge ratio $D/B$. Apparently, the residuals exhibit a tight 
correlation with galaxy $B$-band absolute magnitude but no correlation at all 
with galaxy redshift, mean surface brightness, or disk-to-bulge ratio. 

\begin{figure}[t]
\includegraphics{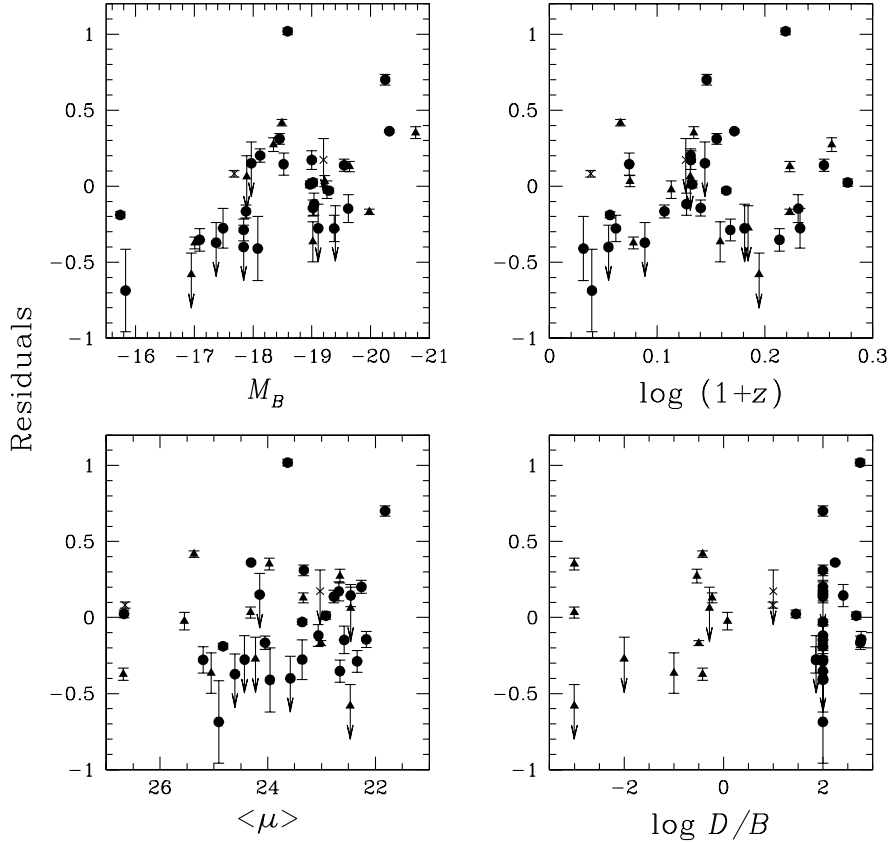}
\vspace{11.5cm}
\caption{Residuals of the $W$ vs. $\rho$ anti-correlation as a function of 
different galaxy parameters. Solid circles represent late-type disk galaxies; 
crosses represent early-type disk galaxies; and triangles represent elliptical 
or S0 galaxies. Arrows indicate $3 \sigma$ upper limits in $W$.}
\end{figure}

Including all the measurable galaxy parameters in the anti-correlation 
analysis, we find that the amount of gas encountered along the line of sight 
depends on galaxy impact parameter and $B$-band luminosity, but does not depend
strongly on galaxy redshift, mean surface brightness, and disk-to-bulge ratio
\cite{c98}. Given the relationship between the gaseous extent $r$ and galaxy 
$B$-band luminosity, we can derive a scaling law which is analogous to the 
Holmberg relation regarding the distribution of luminous matter in galaxies. 
Supplementing results of Chen \etal\ with new measurements, we find
that the gaseous extent of individual galaxies scales with galaxy $B$-band 
luminosity by $r/r_*=(L_B/{L_B}_*)^{t}$ with $t=0.40\pm0.09$ and 
$r_*=204\pm33\ h^{-1}\ {\rm kpc}$ at $W=0.3$ \AA, adopting $q_0=0.5$ and 
$H_0=100\ \kms {\rm Mpc}^{-3}$. 

Based on our analysis, we conclude that extended gaseous envelopes are a 
common and generic feature of galaxies of a wide range of luminosity and 
morphological type. In addition, the result provides for the first 
time a means of quantitatively relating statistical properties of \lya\ 
absorption systems to statistical properties of faint galaxies.

\section{The origin of \lya\ absorption systems}

Given the known gas cross section derived from the galaxy survey, we can 
estimate the incidence of \lya\ absorption lines originating in extended 
gaseous envelopes of galaxies as
\begin{equation}
n(z) = \frac{c}{H_0} (1 + z) (1 + 2 q_0 z)^{-1/2} 
\int^\infty_{L_{B_{\rm min}}} dL_B \ \Phi(L_B,z) \pi r^2(L_B),
\end{equation}
where $c$ is the light speed and $r$ is the gaseous extent scaled with galaxy
luminosity according to the scaling law described in the last section. Adopting
a galaxy luminosity function $\Phi(L_B,z)$ obtained from the Autofib survey at 
redshifts $0.15 < z < 0.35$ \cite{e96}, we find that luminous galaxies can 
explain at least 60\% of the observed number density of \lya\ absorbers 
\cite{w98}.

\begin{figure}[t]
\includegraphics{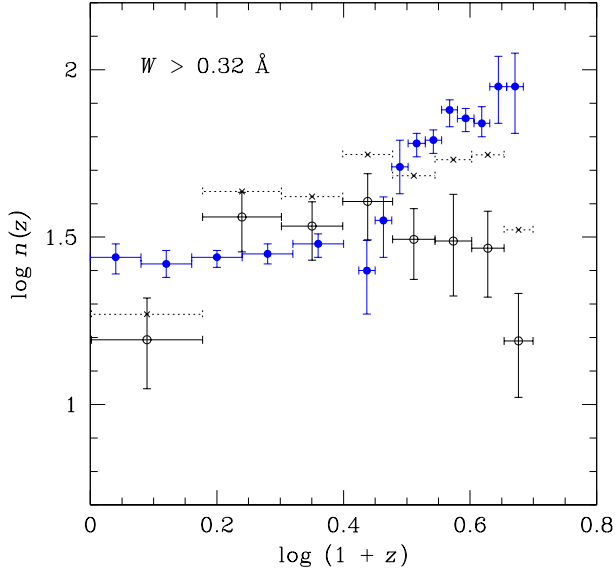}
\vspace{8cm}
\caption{Comparison of the observed number densities of \lya\ absorbers with 
an absorption equivalent width $W>0.32$ \AA\ (closed circles) and the predicted
ones produced in extended gaseous envelopes of galaxies observed in HDF (open 
circles). The crosses with dashed bars indicate the faint-end corrected number 
densities calculated from equation (1) with $L_{B_{\rm min}}=0$.}
\end{figure}

Meanwhile, given that complete and accurate photometric redshift measurements 
are available for galaxies in the Hubble Deep Field (HDF) \cite{f98}, we can 
derive the predicted number density of \lya\ absorbers by multiplying the 
surface density of galaxies, measured empirically from HDF, with the mean gas 
cross section averaged over the sample galaxies, assuming that the gas cross 
section obtained at $z<1$ applies to galaxies at all redshifts with no 
evolution. Equation (1) is now rewritten as
\begin{equation}
n(z) = \frac{S}{(z_2 - z_1)}\sum_{i \atop z_1 \le z_i < z_2} \frac{\pi r^2(L_{B_i})}{D_A^2(z_i)} \Bigg/ N,
\end{equation}
where $(z_1,z_2)$ marks the boundary of each chosen redshift bin, $D_A$ is the
angular distance, $S$ is the surface density of galaxies at redshifts $z_1 <
z < z_2$, and $N$ is the total number of galaxies included in the redshift bin.
The predicted number density of \lya\ absorbers as a function of redshift is 
shown in Fig. 2 together with the observations obtained by Bechtold \cite{b94} 
and Weymann \etal\ \cite{w98}. We also show the predicted number density after 
correcting for galaxies at the faint end of galaxy luminosity function that 
are missing in HDF, adopting a faint-end slope obtained by Ellis \etal\ 
\cite{e96}.  

Clearly, at least 50\% and perhaps as much as 100\% of \lya\ absorption systems
with $W\apg0.32$ \AA\ can be explained by normal galaxies. We conclude that a 
significant amount of \lya\ absorption systems arise in galaxies and that 
\lya\ absorption systems trace galaxies at all redshifts.

\section{Studies of \civ\ absorption systems}

It is generally believed that \civ\ absorption systems arise in the halos of 
intervening galaxies, but the association with galaxies has not been firmly 
established. We repeat the anti-correlation analysis for \civ\ absorbers,
and find that, of all the measurable galaxy properties, the ionized gas cross 
section only depends strongly on galaxy impact parameter and $B$-band 
luminosity. 

\begin{figure}
\includegraphics{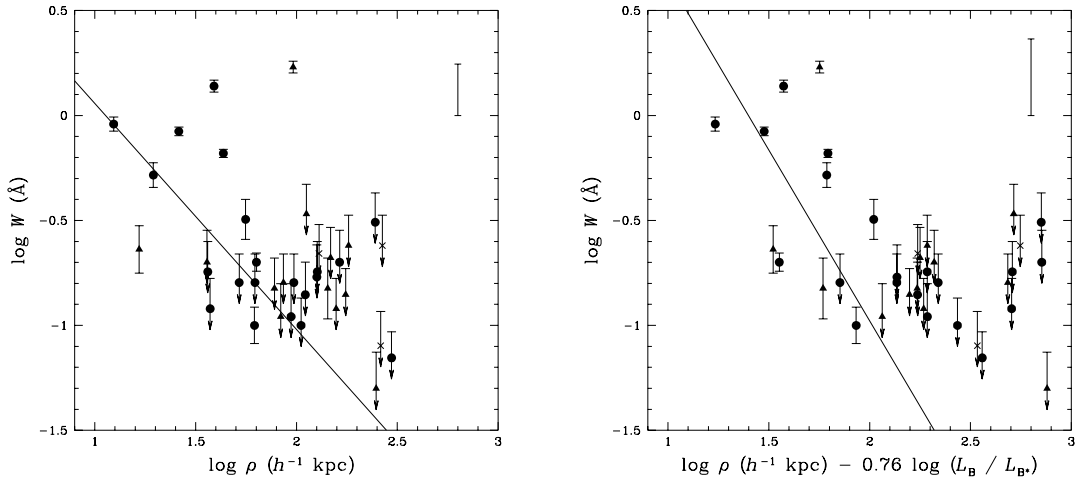}
\vspace{7.0cm}
\caption{Distributions of $\log W$ vs. $\log\rho$ and $\log W$ vs. 
$\log\rho-t\log(L_B/{L_B}_*)$ for \civ\ absorbers. Symbols are the same as 
those in Fig. 1. The cosmic scatter is shown in the corner.}
\end{figure}

We show in Fig. 3 that the \civ\ absorption equivalent width vs. galaxy impact
parameter anti-correlation is significantly improved after accounting for the
galaxy $B$-band luminosity. Consequently we find that the extent of ionized 
gas around galaxies scales with galaxy $B$-band luminosity by 
$r/r_*=(L_B/{L_B}_*)^{t}$ with $t=0.76\pm0.26$ and $r_*=53\pm7\ h^{-1}\ {\rm 
kpc}$ at $W=0.3$ \AA. It is especially interesting to see that the gas cross 
section has a much steeper scaling relation with galaxy luminosity and that the
scaling relation does not evolve strongly with redshift at $z<1$. A more 
extensive analysis will be presented elsewhere.

\acknowledgements{This work was supported by NASA grant NAGW-4422 and NSF 
grant AST-9624216.}


%
\vfill
\end{document}